# Continuum elastic model of fullerenes and the sphericity of the carbon onion shells

Shoaib Ahmad[1,2]

[1]*Accelerator Laboratory, PINSTECH, P.O. Nilore, Islamabad, Pakistan*

[2]*National Centre for Physics, Quaid-i-Azam University, Shahdara Valley, Islamabad 44000, Pakistan*

Email: sahmad.ncp@gmail.com

## Abstract

A continuum elastic model of fullerenes is presented by utilizing the analogy between the closed carbon cages and elastic shells. We derive expressions for the curvature related strain energies $E_p$ of the pentagonal protrusions. We propose to explain the observed sphericity of the carbon onions shells as opposed to the predicted protrusions around the pentagonal defects on the basis of our continuum elastic model of fullerenes. In our model the energy inherent in the pentagonal protrusions $E_p$ is due to the stretching and bending of the shell and shown to be a function of the structural parameters. It also defines the upper limit on the size of the free-standing fullerenes. Using $E_p$ and the topological arguments, we show that the pentagonal protrusions will be smoothed out, resulting in spherical shells of the carbon onions denoted as $C_{60}@C_{240}@C_{540}@C_{960}@C_{1500}$,... .

## I. INTRODUCTION

The energetics and structural stability of closed carbon cages—the fullerenes—has been extensively studied with the binding energy per carbon atom $E_b/n$ being evaluated and used as a structure sensitive as well as a selective parameter for cages with $n$ C atoms. We suggest and show that bending is not the only parameter that is relevant for the cage's elastic behavior; stretching is the predominant effect. We believe that this most pronounced effect responsible for the elastic behavior of fullerene shells has not been fully included in evaluating the curvature-related energies of fullerenes. This article is an attempt to provide a continuum model for the elastic behavior of the spherical shells of fullerenes and the role of the pentagonal protrusions in determining the structure-related properties. We also intend to explain the observed sphericity of the carbon onion shells using this model. Carbon onions have consistently been seen to be composed of spherical shells in electron microscopy [1–4] of soot containing these shelled structures without the 12 protruding regions around the respective pentagons. The repeated observations of spherical onion shells as opposed to the predicted Goldberg polyhedra with protrusions along and around the pentagonal defects [5–7] have warranted an explanation. Some authors [8,9] have provided results from topological arguments to show that the spherical shell structure is more stable. The formation of closed carbon cages in a cluster-forming environment has been extensively studied [10–14]. Delocalization, rehybridization, and electrostatic interactions of the π electrons of the spheroidal cluster atoms provide the stability to these structures [15–17]. In evaluating the binding energies $E_b$ of the $n$ constituents of the fullerenes, the emphasis has generally been on $C_{60}$. Different techniques have been employed such as Hartree–Fock *ab initio* calculations [18], local density approximations (LDA) [19] and tight binding calculations [20]. The binding energy per carbon atom $E_b/n$ is composed of two components $E_b/n = E_{graphene} + E_{cage}$, where $E_{graphene}$ is assumed to be the energy per C atom in graphite's basal plane graphene [21] and $E_{cage}$ is treated as the excess energy related to the curvature of the cage's structure [10,14]. In this article we will evaluate the cage energy in terms of the energy inherent in the pentagonal protrusions $E_p$. Deriving and using the values of $E_p$ combined with the topological arguments, we present a continuum elastic model for the mono and multi-shelled carbon cages.

In calculations of $E_b/n$ the changes and differences in the σ-σ and σ-π angles are included but the structure-related energy is introduced as the bending or the cage energy. Since the elastic properties of shells are very different from those of the plates from the same material, the stretching and bending behavior of hells cannot be directly inferred from the corresponding effects in plates, as was done by some authors [10, 14]. Stretching is the first order and bending the second-order effect in the deformation of shells [22, 23] while the order is reversed in the case of flat plates. In comparing shelled structures with graphene, one must evaluate the energy inherent in shells that is required for conversion into flat plates rather than the other way around. The curvature of shells manifests this behavior, as the strain tensor is inversely proportional to the radius of the shell. The authors of references [10] and [14] obtain radius-independent 'bending' energy of the order of 20 eV for spherical carbon cages with the stretching effect being neglected. We also get a similar result for the bending energy of all fullerenes, but our analysis shows a marked difference, especially when evaluating the pentagonal protrusion energies for higher fullerenes. This extra energy can only be explained when stretching is taken into account. We will show that bending is not the major parameter that is relevant for the carbon cage's elastic behavior; stretching is the predominant effect and should not be neglected.

## II. THE ELASTIC SHELL MODEL OF FULLERENES

In developing our model we clarify that fullerenes may not be directly comparable to a curved piece of graphite's basal plane due to the differences in the elastic behavior of shells versus plates. In the proposed continuum elastic model of fullerenes the order of magnitude estimates of stretching and bending energies are evaluated that a shelled fullerene must possess in order to have the outward protruding curvatures around the 12 pentagons. It is shown that $E_p$ is a structure defining characteristic quantity. An explanation is provided for the absence of the larger fullerenes emerging as stable, independent mono-shelled clusters from the sooting environments. It is due to the higher energy inherent in the defect volume. By using the model to get estimates of the strain energy stored in an outwardly pulled volume which includes the area around the pentagonal protrusions of the fullerene structures, we will provide an explanation for the sphericity of the observed carbon onion shells. To develop an elastic shell model of fullerene cages and to derive the energy $E_p$ inherent in the protrusions, we have made the following assumptions:

(1) The fullerene shell is assumed to be equivalent to the appropriate Goldberg polyhedra with the 12 pentagonal protrusions.
(2) The thickness $t$ of the shell is expected to remain constant during deformation of shape.
(3) The elastic modulus $E$ and the Poisson's ratio $v$ remain constant in fullerenes of different radii and we use the numerical values of these material constants for the basal plane of graphite [21].

In this paper we calculate the cage-related energies as a tool to probe some of the observed elastic properties of fullerenes and the shelled fullerene structures—the carbon onions. We treat $C_{20}$, $C_{60}$, $C_{240}$, $C_{540}$ and the higher fullerenes equivalent to shells of thickness $t$, where $t$ is radius of the cage. The theory of shells provides the orders of magnitude estimates for the pentagonal deformations. Tests of our model would be: (a) identify the role of the curvature around pentagons; (b) provide the reasons for unusual stability of $C_{60}$; (c) set the approximate limit on the size of the largest free-standing mono-shelled fullerene in a sooting environment; and (d) explain the sphericity of the carbon onions by suggesting possible routes to the elimination of the predicted pentagonal protrusions in these multi-shelled carbon cages.

### A. Deformation energies of shells versus tubules and plates

Existence of spherically curved surfaces introduces additional forces that depend on the properties of the particular surfaces. In the case of the separation of phases, the thermodynamic properties are determined by the coefficient of surface tension $T$, which is related to the pressure difference $\Delta P$ across the interface according to Laplace's equation $\Delta P = 2T/R$, where $R$ is the radius of curvature. Elasticity theory relates $T$ with the tangential stresses, and the equilibrium equation for a spherical shell involves these stress tensors. Analyses that have evaluated bending strain energies usually rely on the method developed by Tibbetts [24]. His analysis is based on the bending of a rectangular plate into a cylinder. Conversion of a plate into a tubule does not require the stretching effect to be considered unless the radius of tubule $R$ approaches thickness $t$. Such large deformations of planar structures involve stretching as a second-order effect, while bending is the first order effect in such transformations. Therefore, the strain energies obtained for tubules with the considerations of bending only provide fairly accurate estimates, whereas, for smaller $R$ of the order of $t$, a correction due to stretching can be included in the tubular deformation energy. On the other hand, the deformation energies required for the introduction of protrusions in any shelled structure involve the stretching and bending effects, in that order. The complete neglect of a significant first-order stretching effect may have been responsible for the fairly low cage energies per C in most analyses, especially Refs. [10] and [14]. Stretching has been included along with bending in Refs. [25–27] in order to evaluate the curvature-related energies of fullerenes. However, the protrusions that are so pronounced in all structures, except in the case of $C_{60}$, have not been treated in these models.

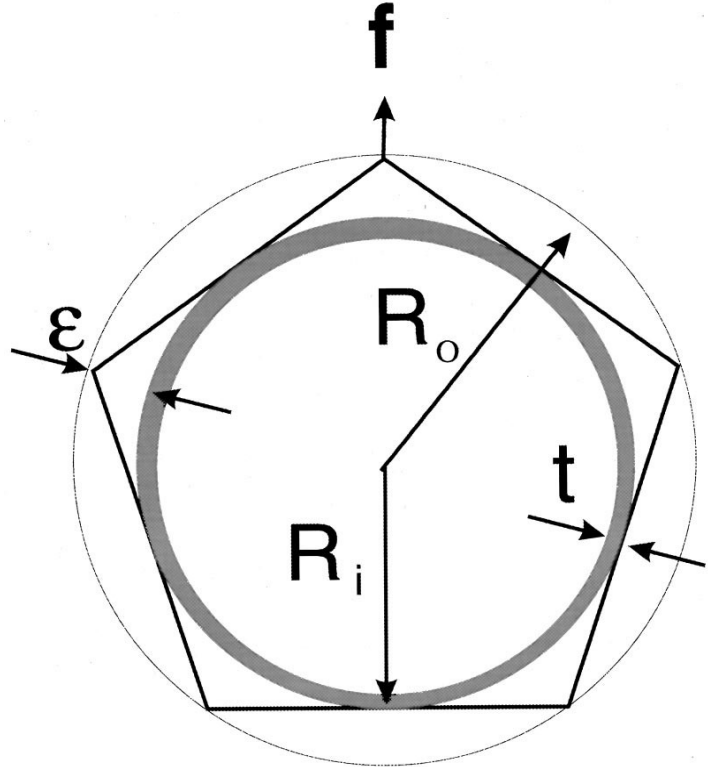

Figure 1. A pentagonal protrusion is superimposed on a sphere of radius $R_i$. An outward normal force $f$ is being applied at 12 symmetrically disposed protruding points to convert a sphere into a Goldberg polyhedron with circumscribing sphere of radius $R_o$. In fullerenes the radial extension $\zeta$ is equivalent to $\varepsilon$.

## B. Estimates of stretching $E_{str}$ and bending $E_{ben}$ energies of shells

We estimate the energy required to stretch a spherical shell of radius $R$ and thickness $t$ into a structure with 12 pentagonal protrusions with the inscribing radius $R_i$ and circumscribing radius $R_o$ as shown in Figure 1. The shell undergoes uniform stretching with radial extension $\zeta$. The length of equatorial circumference increases by $2\pi\zeta$ with the relative extension being $\zeta/R$, as the strain tensor $u_{str} \propto \zeta$ and the corresponding stress tensor $\sigma_{str} \sim E\zeta/R$. The stretching energy per unit area is given in an order of magnitude estimate as $E_{str} \sim Et(\zeta/R)^2$. Similarly, the bending energy $E_{ben}$ associated with the deformation of shells can be estimated as $E_{ben} \sim Et^3(\zeta/R)^2$.
A radius-independent $E_{ben} \approx 20$ eV for all closed caged carbon structures can be obtained by assuming the shell of radius $R$ to be composed of two equal halves of surface area $2\pi R$. If we neglect stretching, then the bending of one of the half-shells can convert it into a spherical sheet of thickness $t$. In this case $\zeta = 0.414R$. The bending energy for the two half shells is $E_{ben} \sim Et^3(\zeta/R)^2 = 0.17\pi Et^3 \approx 20\ eV$ by using the values of $E$ and $t$ given in Ref. [21]. Thus, we obtain a radius-independent bending energy as obtained by other authors [10–14].

## C. Energy required for the pentagonal protrusions

Let a spherical shell of thickness $t$ be subjected to a concentrated force $f$ along the outward normal as shown in Figure 1. The resulting deformation is $\zeta = R_o - R_i$, where $R_i$ and $R_o$ are the radii of the inscribing and circumscribing spheres, respectively. $\zeta$ will vary over a perpendicular distance $d$. The inscribing shell's radius $R_i$ can be considered to be equivalent to the radius of the corresponding fullerene. The distance $d$ is a measure of the region of outward protrusion and the area of deformation $\sim d^2$. The bending energy over this area is $E_{ben} \sim Et^3(\zeta/d)^2$, while the respective stretching energy $E_{str} \sim Et[(\zeta d)/R]^2$. The bending energy decreases and stretching increases with the increase in $d$, thus, both energies should be considered in determining the deformation near the point of application of the force. Minimizing their sum ($E_{ben} + E_{str}$) with respect to $d$, one gets

$$d \sim \sqrt{Rt} \qquad (1)$$

The total deformation energy of a pentagonal protrusion $E_p = E_{ben} + E_{str}$ with $d \sim \sqrt{Rt}$ yields

$$E_p \sim 2Et^2(\zeta^2/R) \qquad (2)$$

$E_p$ is the net amount of energy required to stretch a portion of an area $\sim d^2$ to protrude a volume $\delta V \approx \zeta d^2$ outward by the application of a concentrated, normal, and outwardly directed force $f$. The resulting amount of deformation $\zeta$ is not small compared with the radius. The total defect energy of the resulting shell is the so-called cage energy $E_{cage} \equiv E_p^{tot} = 12 \times E_p$ for the 12 pentagonal protrusions. A somewhat similar and interesting qualitative analysis has been done to study the compressions of thin, i.e., $R \gg t$ elastic shells [29] upon contact with flat surfaces. The differences in that model and ours are inherent in the basic assumptions of the tensile versus compressive deformations. Our method is based on the outward protruding defect regions of thick ($t \le R$) shells with $R\zeta \ge d^2$, while their results are mostly valid for thin shells ($t \ll R$) and with the condition that $R\zeta \ll d^2$.

## D. Cage energy of $C_{60}$

The 12 pentagons are uniformly shared on the perfect spherical surface of $C_{60}$ and their deformation energies are uniformly distributed over the entire shell. We do not observe the protrusions that are expected in all larger fullerenes. To estimate the cage energy inherent in $C_{60}$'s shell structure, we use Eq. (2) for evaluating the energy required to deform a half-shell into a circular plate without the change in thickness. For such a large deformation $\zeta = 0.414R$, where $R$ is the radius of the shell. The total cage energy of the two half-shells is given by

$$E(C_{60}) \sim 4Et^2\zeta^2/R(C_{60}) = 0.68Et^2R(C_{60}) \approx 50 \text{ eV}.$$

## III. DISCUSSION

One of the main objectives of the paper is to clarify that fullerenes may not be directly comparable to a curved piece of graphene. In presenting a continuum elastic shell model of fullerenes, we have evaluated the orders of magnitude estimates of the deformation energies of the pentagons that cause the curvature in an otherwise flat piece of graphene. Shells have almost an order of magnitude higher structure-related energy compared with the tubules of comparable radii. This fact may play a decisive role in the energetics of the formation of fullerenes in sooting environments. On the basis of these results the estimates are obtained for the elastic energy stored in a volume $\Delta V$, which includes the adjacent areas around the pentagonal protrusions of the fullerene structures. We notice that considerably large amounts of energy are stored within these polyhedral regions. These strain energies are shared by the entire shell.

We have evaluated the cage energies per carbon atom $E_{cage}/n$ obtained from the pentagonal protrusion energies $E_p$ by using Eq. (2) for the successive Goldberg polyhedra. A comparison with $E_{ben}/n$ is also provided to highlight the essential difference in our treatment of this problem. Although we have started from $C_{20}$ as the first shell which has 20 trigonal protrusions, the model is however equally valid for $C_{20}$. Similarly, in the case of $C_{60}$ we do not have the pentagonal protrusions as in larger fullerenes; therefore, the strain energy is evaluated from the deformation energy required to deform a shell into two flat circular disks. The results of this strain energy are shown in the same figure with $C_{20}$ and the higher fullerenes. The energy required for producing a localized protrusion becomes larger than that required for shell's bending energy for all fullerenes $\geq C_{540}$; the individual protrusion has $E_p \geq E_{ben} \sim 20$ eV. $E_p$ is a gradually increasing function of the shell radius. Therefore, the authors that assume bending only as the cage energy for all fullerenes ($E_{cage}=E_{ben}\approx 20$ eV) have obtained a rapidly decreasing deformation energy per carbon atom $E_{cage}/n$ for higher fullerene shells.

We have also shown $E_{ben}/n=20/n$, where $n$ is the number of C atoms in respective shells.

| $C_n$ | $R_i$ [Å] | $R_0$ [Å] | $\zeta$ [Å] | $E_p$ [eV] | $\Sigma d^2/4\pi R_i$ | $E_{cage}/n$ [eV C$^{-1}$] | $E_{ben}$ [eV C$^{-1}$] |
|---|---|---|---|---|---|---|---|
| $C_{20}$ | 1.95 | 2.4 | 0.46 | 3.5 | 0.9 | 3.5 | 1.0 |
| $C_{60}$ | 3.52 | … | … | … | 0.33 | 0.83 | 0.33 |
| $C_{280}$ | 7.05 | 8.71 | 1.66 | 16.2 | 0.25 | 0.81 | 0.083 |
| $C_{540}$ | 10.54 | 13.03 | 2.49 | 24.3 | 0.17 | 0.54 | 0.037 |
| $C_{960}$ | 14.04 | 17.35 | 3.31 | 32.3 | 0.13 | 0.4 | 0.021 |
| $C_{1500}$ | 17.53 | 21.67 | 4.14 | 40.5 | 0.098 | 0.32 | 0.013 |
| $C_{2160}$ | 21.03 | 25.99 | 4.96 | 48.4 | 0.08 | 0.27 | 0.009 |

TABLE I. The results are shown for the first few fullerenes with radii calculated from Ref. [28] except $C_{60}$. The table presents the results of the calculations of $E_p$, the defect energies of the predicted Goldberg polyhedral protrusions from Eq.(2). The radii of the circumscribing $R_o$ and inscribing $R_i$ spherical surfaces are calculated using the Golden ratio. $\Sigma d^2/4\pi R_i^2$ is the ratio of the total defect to the surface areas. In the last two columns the cage energy of the carbon atoms in successive cages is calculated as $E_{cage}/n$ [eVC$^{-1}$] and $E_{ben}/n$ [eVC$^{-1}$]. We have used $E_{ben} \sim 20$ eV.

These values are tabulated in Table I. Table I is presented to understand the curvature-related energies of the carbon cages and to develop topological arguments for the sphericity of the carbon onion shells. Essential features of the model are presented for the first few members of fullerene spheroids—the Goldberg polyhedra5 with vertices given by $n = 20(b^2 + bc + c^2)$, where $b=c=1, 2, 3$, for $C_{60}, C_{240}, C_{540}$, respectively. In the case of $C_{20}$, $b=1$ and $c=0$. The radii are calculated from Ref. [25], thickness $t=1.82$ Å, i.e., the area of C atoms by using $\rho=4.16$ [g cm$^{-3}$] for graphene, the modulus of elasticity $E=10^3$ GPa for the basal plane and the Poisson ratio $\nu = 0.163$ [21]. We have tabulated the radii of the circumscribing ($R_o$) and inscribing ($R_i$) spherical surfaces. The circumscribing sphere's radius is evaluated for the pentagon enclosed within by using the Golden section $\tau=1.618$. Stability of $C_{60}$ is due to the low cage energy per C in the pentagonal defect regions of the shell compared with other cages considered. For $C_{60}$ the $E_p/C\approx 0.81$ eV per C compared with a value of 3.3 [eVC$^{-1}$] for all other structures with protrusions.

Topological arguments in favor of spherical shells come from the mismatch between the radii of the inscribing sphere of the larger shell and the circumscribing sphere of the smaller one. For example, $R_o$ of $C_{1500}$ is 21.67 Å and it is $R_i$ of $C_{2160}$. Therefore, $C_{1500}$ cannot be contained within $C_{2160}$. In fact, no onion configuration with protrusions will provide inter-shell separation of 3.34 Å as is seen in Refs. [2–4]. Thus, all shells are compressed by their successive higher neighbors. This geometrical requirement introduces sphericity, and onions with larger shells become energetically favored due to the elimination of protrusions.

## IV. CONCLUSIONS

We have drawn a close analogy between fullerenes and the elastic shells of finite thickness. Mostly the single-shelled spherical structures are discussed but the multi-shelled carbon onions have also been considered, especially on the issue regarding the absence of the predicted Goldberg polyhedral with pronounced pentagonal corrannulene regions extending outwards. These protrusions must have been observed along C5 axis of symmetry. The high energy and high intensity electron irradiation of soot1–3 and the onions produced by energetic C ion implantations in Cu,4 have confirmed the emergence of spheroidal onion shells in TEM, HREM, and AFM. It has been observed that in any given onion configuration denoted as $C_{60}@C_{240}@C_{540}@C_{960}@C_{1500}$,…., none of the component fullerenes shells shows the protruding areas around the 12 pentagons symmetrically disposed around the surface.

The significant differences in the three carbon cluster structures—the graphene sheets, tubules, and the shelled fullerenes—have been pointed out. Whereas the elastic strain energy for the tubules can be worked out from the bending of rectangular or triangular graphene plates, the end capping of the open end of a cylindrical tubule requires six pentagons and a certain number of hexagons. The presence of these end caps, whose radius is determined by the radius of the tubule, requires much larger energy per carbon than is needed for the adjoining cylindrical structure. Shells are shown to possess much higher energy $E_p$ for all closed carbon cages $C_n$ due to the inherent curvature due to the pentagonal defect protrusions. Since these pentagons do not protrude in the carbon onion structures, therefore the defect's strain energy may be uniformly spread over the entire surface, resulting in perfectly spherical carbon onion shells. The shells' protrusions have $Ep \propto \zeta^2 R^{-1}$. The model also shows that $E_{\text{ben}} \approx 20$ eV for all shells regardless of the size of their radii.

We would like to point out the explanation of the absence of much larger fullerenes emerging from a sooting environment as independent and stable mono-shelled carbon cages due to the high strain energy inherent in the defect volume. The limit on the largest free-standing fullerenes may eventually be set by the large $E_p$ required for fullerenes $\geq C_{540}$. This may be considered to be the justification that the semi-prophetic prediction of Daedalus of the ''large graphitic balloons'' dreamt in a 'charcoal fire' [30] cannot be fulfilled since shells with larger radii require more defect energy constrained within each protrusion than that required to transform a spherical shell into a circular sheet to, i.e., $E_{\text{ben}}$.

## V. REFERENCES

[1] S. Ijima, J. Cryst. Growth **50**, 657, 1980; J. Phys. Chem. **91**, 3466,1987.
[2] D. Ugarte, Nature (London) **359**, 707, 1992; Europhys. Lett. **22**, 45, 1993.
[3] M. S. Zwanger and F. Banhart, Philos. Mag. B **72**, 149, 1995.
[4] T. Cabioch, J. P. Riviere, and J. Delafond, J. Mater. Sci. **30**, 4787, 1995.
[5] P. W. Fowler, Chem. Phys. Lett. **131**, 444, 1986; P. W. Fowler and D. E. Manolopoulous, *An Atlas of Fullerenes,* Clarendon, Oxford, 1995.
[6] H. W. Kroto, Science **242**, 1139,1988; Nature (London) **670**, 359, 1992.
[7] A. Maiti, C. J. Brabec, and J. Bernholc, Phys. Rev. Lett. **70**, 3023, 1993.
[8] J. P. Lu and W. Yang, Phys. Rev. B **49**, 11421, 1994.
[9] E. Pasqualini, Phys. Rev. B **56**, 7751, 1997.
[10] T. G. Schmalz, W. A. Seitz, D. J. Klein, and G. E. Hite, J. Am. Chem. Soc. **110**, 1113, 1988.
[11] D. Bakowies and W. Thiel, J. Am. Chem. Soc. **113**, 3704, 1991.
[12] D. Tomanek and M. A. Schluter, Phys. Rev. Lett. **67**, 2331, 1991!.
[13] G. B. Adams, O. F. Sankey, J. B. Page, M. O'Keeffe, and D. A. Drabold, Science **256**, 1792, 1992.
[14] K. H. Bennemann, D. Richardt, J. L. Moran-Lopez, R. Kerner, and K. Penson, Z. Phys. D: At., Mol. Clusters **29**, 231, 1994.
[15] P. W. Fowler and J. Woolrich, Chem. Phys. Lett. **127**, 78, 1986.
[16] R. C. Haddon, L. E. Brus, and K. Ranghavachari, Chem. Phys. Lett. **131**, 165, 1986.
[17] D. P. Clougherty and X. Zhu, Phys. Rev. A **56**, 632, 1997.
[18] J. Husak and H. Schwarz, Chem. Phys. Lett. **205**, 187, 1993.
[19] S. Saito and A. Oshiyama, Phys. Rev. Lett. **66**, 2637, 1991.
[20] B. L. Zhang, C. H. Xu, C. Z. Wang, C. T. Chan, and K. M. Ho, Phys. Rev. B **46**, 7333, 1992.
[21] B. T. Kelly, *Physics of Graphite*, Applied Sciences, London, 1981, Chap. 2.
[22] L. D. Landau and E. Lifshitz, *Theory of Elasticity*, 3rd ed. Pergamon, London, 1986, Chap. 3.
[23] S. Timoshenko and S. W. Krieger, *Theory of Plates and Shells*, 2nd ed. ,McGraw-Hill, New York, 1959, Chaps. 13 and 14.
[24] G. G. Tibbetts, J. Cryst. Growth **66**, 632, 1984.
[25] T. Tersoff, Phys. Rev. B **46**, 15546, 1992.
[26] T. Whitten and H. Li, Europhys. Lett. **23**, 5, 1993.
[27] X. Xiaoyu, L. Ji-xing, and O. Y. Zhong-can, Mod. Phys. Lett. B **9**, 1649, 1994.
[28] M. Yoshida and E. Osawa, Fullerene Sci. Technol. **1**, 55, 1993.

[29] L. Pauchard and S. Rica, Philos. Mag. **78**, 225, 1998.
[30] D. E. H. Jones, New Sci. **35**, 245, 1966; *The Fullerenes*, edited by H. W. Kroto, D. R. M. Walton, Cambridge University Press, Cambridge, 1993, pp. 9–18.